\documentclass[floats,aps,showpacs,superscriptaddress,amsmath,amssymb,prb,twocolumn]{revtex4-1}
\usepackage{graphicx}
\usepackage{color}
\usepackage{amsmath}
\usepackage{rotating}
\usepackage{dcolumn}
\usepackage{bm}

\begin{document}
\newcommand{\g}{\bf}

\title{Evolution of the persistent spin helix in the presence of  Hartree-Fock fields}

\author{Matthias C. L\"uffe}
\affiliation{\mbox{Dahlem Center for Complex Quantum Systems and Fachbereich Physik, Freie Universit\"at
  Berlin, 14195 Berlin, Germany}}
 \author{Jeroen Danon}
 \affiliation{\mbox{Dahlem Center for Complex Quantum Systems and Fachbereich Physik, Freie Universit\"at
  Berlin, 14195 Berlin, Germany}}
\affiliation{\mbox{Niels Bohr International Academy, Niels Bohr Institute, Blegdamsvej 17, DK-2100 Copenhagen, Denmark}}
\author{Tamara S.\ Nunner}
\affiliation{\mbox{Dahlem Center for Complex Quantum Systems and Fachbereich Physik, Freie Universit\"at
  Berlin, 14195 Berlin, Germany}}

\date{\today}

\begin{abstract}
We derive a spin diffusion equation for a spin-orbit coupled two-dimensional electron gas including the Hartree-Fock field resulting from 1st order electron-electron interactions. We find that the lifetime of the persistent spin helix, which emerges for equal linear Rashba- and Dresselhaus spin-orbit interactions, can be enhanced considerably for large initial spin polarizations due to the Hartree-Fock field. The reason is a reduction of the symmetry-breaking cubic Dresselhaus scattering rate by the Hartree-Fock field. Also higher harmonics are generated and the polarization of the persistent spin helix rotates out of the ($S_y$,$S_z$)-plane acquiring a finite $S_x$-component. This effect becomes more pronounced, when the cubic Dresselhaus spin-orbit interaction is large.
\end{abstract}

 \maketitle

\section{Introduction}

In recent years the spin dynamics in semiconductors has attracted great interest, because integrating the spin degree of freedom into semiconducting devices opens up new possibilities of realizing spintronics or quantum computational devices.~\cite{Zutic} For these potential applications long spin lifetimes are highly desirable. In this respect the persistent spin helix (PSH),~\cite{schliemann,bernevig1} a spin-density wave state with ideally infinite lifetime, is of great interest. This state exists in two-dimensional electron systems with Rashba and linear Dresselhaus spin orbit interactions of equal magnitude due to a SU(2) symmetry of the underlying Hamiltonian.~\cite{bernevig1} On a less abstract level, the PSH state emerges because for Rashba and linear Dresselhaus spin orbit interactions of equal magnitude the resulting spin orbit interaction points into the same direction for all electronic momentum states. The magnitude of the spin-orbit interaction grows linearly with the projection of the momentum on the PSH wave-vector, such, that all spins rotate by the same angles when they diffuse along the PSH direction.

In real systems the lifetime of the PSH is no longer infinite due to the presence of additional terms which break the SU(2) symmetry of the Hamiltonian. It has been suggested, that the presence of cubic Dresselhaus spin orbit interaction is the dominant symmetry-breaking mechanism leading to a finite lifetime of the PSH state.~\cite{koralek,YangTheory,Matthias} Experimentally, the PSH state has been observed in GaAs/AlGaAs quantum wells,~\cite{koralek,Yang,Walser,YangNew} where the cubic Dresselhaus spin orbit interaction is quite small, as well as in (In,Ga)As quantum wells,~\cite{Kohda} where the cubic Dresselhaus spin orbit interaction is much larger. 

The observed temperature dependence of the lifetime of the PSH state~\cite{koralek} suggests, that also electron-electron interactions strongly affect the lifetime of the PSH state.~\cite{Matthias,Sinova} In our previous work~\cite{Matthias} we found that the consideration of electron-electron interaction in combination with cubic Dresselhaus spin orbit interaction as the symmetry breaking mechanism can fairly well account for the observed temperature dependence of the PSH lifetime. In this previous treatment we have only considered electron-electron collisions, which arise in 2nd order in the electron-electron interaction but we have neglected the Hartree-Fock term, which arises in 1st order. This approximation is valid, when the initial spin polarization is small and seems to be appropriate for the present experimental observations of the PSH state.~\cite{koralek} On the other hand, it is experimentally also possible to realize large initial spin polarizations,~\cite{Stich} where the Hartree-Fock term, which acts like an effective magnetic field pointing along the local spin polarization, is known to enhance the spin lifetime considerably.~\cite{Stich,WengWu,Takahashi} Therefore it is the purpose of the present paper to analyze the effect of the Hartree-Fock term on the PSH in the diffusive D'yakonov Perel' regime. In particular we want to answer the following questions: (i) what is the effect the Hartree-Fock term on the lifetime of the PSH-state and (ii) does it modify the pattern of the spin-polarization of the PSH-state?

This article is organized as follows: In Sec.~\ref{sec:SpinDiffusionEquation} we present the spin diffusion equation for a two-dimensional electron gas with Rashba- and Dresselhaus spin-orbit interactions including the Hartree-Fock field, which arises from 1st order electron-electron interaction. In Sec.~\ref{sec:Results} we show how the presence of the Hartree-Fock term enhances the lifetime of the PSH state. Furthermore we demonstrate that the presence of the Hartree-Fock term also generates higher harmonics in the spin polarization. In particular, we show that the 2nd harmonic corresponds to an $S_x$-component and thus rotates the spin-polarization out of the $(S_y,S_z)$-plane. The appendices contain details about the derivation of the Hartree-Fock term (appendix~\ref{app:HartreeFock}) and of the spin diffusion equation (appendix~\ref{app:SpinDiffusionEquation}).

\section{Model}
\label{sec:Model}

We describe conduction band electrons in a (001) grown quantum well with Rashba- and Dresselhaus spin-orbit interactions, potential impurities and electron-electron interactions. For simplicity we focus on the limit of zero temperature, where the 2nd order electron-electron scattering vanishes but the 1st order contribution due to electron-electron interaction, i.e., the Hartree-Fock term, is finite. Although our treatment is strictly valid only at zero temperature, we expect, that qualitatively similar results hold also for small but finite temperatures, because the temperature dependence of the Hartree-Fock term is weak and the presence of 2nd order electron-electron interactions does not qualitatively affect the PSH state.~\cite{Matthias}

We consider the following Hamiltonian (here, and in the following use $\hbar=1$):
\begin{equation}
H = H_\textnormal{0}+H_\textnormal{imp}+H_\textnormal{e-e}
\label{eq:H},
\end{equation}
where the first term represents a two-dimensional electron gas (2DEG) 
with a quadratic dispersion and intrinsic spin-orbit interaction
\begin{equation}
H_0 = \sum_{s,s'; \bf{k}} \psi_{\bf{k} s'}^\dagger\,\mathcal{H}_{0 s's}\,\psi_{\bf{k} s}
\label{eq:general2ndq}
\end{equation}
and the $2 \times 2$ matrix in spin space
\begin{equation}
\mathcal{H}_0 = \epsilon_k+ \bf{b}(\bf{k})\cdot {\boldsymbol{\sigma}}.
\label{eq:H0}
\end{equation}
 The $\psi_{\bf{k} s}^\dagger \left(\psi_{\bf{k} s}\right)$ are creation (annihilation) operators for electrons with momentum $\bf{k}$ and spin projection $s$. Within the standard envelope function approximation\cite{winkler} one finds  $\epsilon_k=\frac{k^2}{2\, m}$ where $m$ is the effective mass. The vector of Pauli matrices is denoted by $\boldsymbol{\sigma}$  and the in-plane spin-orbit field
\begin{equation}
{\bf b}({\bf k}) = {\bf b}_R({\bf k})+{\bf b}_{D}({\bf k})+{\bf b}_{cD}({\bf k})
\label{eq:bk}
\end{equation}
contains Rashba spin-orbit interaction
\begin{equation}
{\bf b}_R({\bf k}) = \alpha v_F
\begin{pmatrix}
k_y\\ -k_x \\ 0
\end{pmatrix}
\label{eq:RashbaSOI}
\end{equation}
as well as linear and cubic Dresselhaus spin-orbit interactions
\begin{equation}
{\bf b}_{D}({\bf k}) = \beta v_F
\begin{pmatrix}
k_y\\ k_x \\ 0
\end{pmatrix}, \,\,\,\,\,
{\bf b}_{cD}({\bf k}) =
\gamma v_F \frac{k^3}{4}
\begin{pmatrix}
\sin 3 \theta \\
-\cos 3 \theta \\0
\end{pmatrix}.
\label{eq:linearDresselhausSOI}
\end{equation}
Here, $v_F$ is the Fermi velocity, the angle $\theta$ gives the direction of $\bf{k}$ with respect to the $x$ axis, which we choose to point along the (110)-crystal direction.

Furthermore, we have included in Eq.~(\ref{eq:H}) scalar electron-impurity interactions,
\begin{equation}
H_\textnormal{imp} = \frac{1}{V} \sum_{s;{\bf k}, {\bf k'}}\psi_{{\bf k'} s}^\dagger U_{{\bf k' k}}\, \psi_{{\bf k}s},
\label{eq:Himp}
\end{equation}
arising from scattering of electrons at potential impurities $V^\textnormal{imp}({\bf x})=\sum_i v({\bf x}-{\bf R}_i)$, where $v({\bf x})$ denotes the potential of each single impurity and $\left\{ {\bf R}_i\right\}$ are the impurity positions, eventually to be averaged over. 

Finally, the Hamiltonian~(\ref{eq:H}) contains electron-electron interactions, 
\begin{equation}
H_\textnormal{e-e} = \frac{1}{2V} \sum_{\substack{{\bf k_1}\dots {\bf k_{4}}\\ s_1, s_2}}V_{{\bf k_{3}},{\bf k_{4}},{\bf k_{1}},{\bf k_{2}}}\,\psi_{{\bf k_{4}} s_2}^\dagger \psi_{{\bf k_{3}} s_1}^\dagger \psi_{{\bf k_1} s_1}\psi_{{\bf k_2} s_2}
\label{eq:Hee}
\end{equation}
with a Thomas-Fermi screened Coulomb potential of the form $V_{{\bf k}_{3},{\bf k}_{4},{\bf k}_{1},{\bf k}_{2}}\approx \tilde v(|{\bf k}_{3}-{\bf k}_{1}|) \delta_{{\bf k}_1+{\bf k}_2-{\bf k}_3-{\bf k}_4,0}$ where $\tilde v(q)=\frac{2\,\pi}{\varepsilon_q m\, q\, a^*}$ and $\varepsilon_q \approx1+\frac{2}{q\, a^*}$ with $a^*=\frac{4 \,\pi\, \varepsilon_0\,\varepsilon_r }{m\, e^2}$ being the effective Bohr radius.

Following the procedure of our previous treatment of the PSH-state\cite{Matthias} we use the Nonequilibrium statistical operator method\cite{zubarev} to derive kinetic equations for the charge and spin components of the Wigner-transformed density matrix
\begin{equation}
\hat{\rho}_{\bf k}({\bf x},t) = n_{\bf k}({\bf x},t)+{\bf s}_{\bf k}({\bf x},t) \cdot {\boldsymbol{\sigma}}.
\end{equation}
Since, to zeroth order in ${b}/E_F$, the kinetic equations for spin and charge decouple we only need to consider the spin component and find for the spin density ${\bf s}_{\bf k}$:
\begin{eqnarray}
\label{eq:ChargeSpin}
\partial_t\,{\bf s}_{\bf k}+{\bf v}\cdot \partial_{\bf x} {\bf s}_{\bf k} +2\,{\bf s}_{\bf k}\times{\bf b}({\bf k}) &=& {\bf J}^\textnormal{imp}_{\bf k} +{\bf J}^\textnormal{ee(1)}_{\bf k} 
\end{eqnarray}
with $v_i=k_i/m$, where the index $i=x,y$ labels the in-plane spatial directions. For simplicity, we focus here on the limit of zero temperature, where the 2nd order electron-electron collision integral vanishes and we keep only the first order electron-electron collision integral ${\bf J}_{\bf k}^{\rm ee(1)}$.  As we show in appendix~\ref{app:HartreeFock}, ${\bf J}_{\bf k}^{\rm ee(1)}$ can be cast into an effective magnetic field, i.e., a Hartree-Fock term, arising from the mean polarization of the PSH-state itself
\begin{equation}
{\bf J}_{\bf k}^{\rm ee(1)} = - 2 \, {\bf s}_{\bf k} \times \int \frac{d {\bf q}}{(2 \pi)^2} \tilde v(q) {\bf s}_{\bf k+q}\,.
\label{eq:1storderEECollisionIntegralmt}
\end{equation}
The impurity collision integral in Eq.~(\ref{eq:ChargeSpin}) is given by:
\begin{eqnarray}
{\bf J}^\textnormal{imp}_{\bf k} &=& -\sum_{\bf k'} W_{\bf k k'} \delta(\epsilon_{\bf k}-\epsilon_{\bf k'}) ({\bf s}_{\bf k} - {\bf s}_{\bf k'})\,,
\end{eqnarray}
where the transition rate is given by $W_{\bf k k'}= {2 \pi} n_i |v \left({\bf k'}-{\bf k}\right)|^2$ with the impurity concentration $n_i$.

\section{Spin-diffusion equation including Hartree-Fock fields}
\label{sec:SpinDiffusionEquation}

In the diffusive strong scattering D'yakonov-Perel' regime $b (k_F) \tau \ll 1$ (where $\tau$ is the momentum relaxation time) one can derive a diffusion equation for the spin density of the form (see appendix~\ref{app:SpinDiffusionEquation} for details)
\begin{eqnarray}
\partial_t \bf{S} &=&
\hat D {\bf S} +{\bf H}_1 + {\bf H}_3 \,.
\label{eq:SpinDiffusionEquation}
\end{eqnarray}
We focus for simplicity on equal Rashba and Dresselhaus spin-orbit interactions $\alpha=\beta$, i.e., the usual condition for the PSH state. In this case we find for the matrix $\hat D$
\begin{equation}
\hat D=
\begin{pmatrix} \!
{\tilde D} \partial_y^2\!-\!{\tilde \gamma}_{\rm cd} \!\!\!&\!\!\! 0 \!\!\!&\!\!\! 0 \\
0 \!\!\!&\!\!\! {\tilde D}( \partial_y^2 \!-\!q_0^2 ) \!-\!{\tilde \gamma}_{\rm cd} \!\!\!&\!\!\! 2 q_0 \tilde D \partial_y \\
0 \!\!\!&\!\!\! -2 q_0 \tilde D \partial_y \!\!\!&\!\!\! {\tilde D} (\partial_y^2 \!-\!q_0^2 ) \!-\! 2 {\tilde \gamma}_{\rm cd}
\end{pmatrix}
\label{eq:hatD}
\end{equation}
where $q_0=4 m v_F \alpha$ is the PSH wave-vector. This corresponds to the usual spin diffusion matrix in the absence of Hartree-Fock fields but with a renormalized diffusion constant $\tilde D$ and a renormalized cubic Dresselhaus scattering rate $\tilde \gamma_{\rm cd}$, given by
\begin{equation}
\tilde D = \frac{D}{1+(2 B_1 \tau)^2}, \quad
\tilde \gamma_{\rm cd} = \frac{\gamma_{\rm cd}}{1+(2 B_3 \tau)^2} \,,
\label{eq:tildeGamma}
\end{equation}
where $D=\frac{1}{2}v_F^2 \tau$ is the ordinary diffusion constant and $\gamma_{\rm cd}=\frac{1}{8}v_F^2 \gamma^2 k_F^6 \tau$ is the cubic Dresselhaus scattering rate.
We also introduced the effective Hartree-Fock fields
\begin{equation}
{\bf B}_{1,3}= \chi_{1,3} {\bf S},
\label{eq:HartreeFockField}
\end{equation}
which are the Hartree-Fock fields for the winding number one (${\bf s}_{\pm 1}$) and winding number three (${\bf s}_{\pm 3}$) components of the spin density
\begin{equation}
{\bf s_k}=\sum_{l=0,\pm1,\pm 3} {\bf s}_l(k) e^{i l \theta}.
\end{equation}
For $\chi_{1,3}$ we find (see appendix~\ref{app:HartreeFock} and appendix~\ref{app:SpinDiffusionEquation})
\begin{equation}
\chi_{{\bf k},n} = \frac{4 \pi}{m} \int \frac{d {\bf q}}{(2 \pi)^2} \tilde v(|{\bf q}-{\bf k}|) f'(\epsilon_q) \left( 1- \frac{q^n}{k^n} \cos (n \theta_{\bf qk}) \right),
\label{eq:Chi13}
\end{equation}
where $\theta_{\bf qk}$ is the angle between ${\bf k}$ and ${\bf q}$. Assuming a Fermi temperature of $T_F=400K$, an effective mass of $m_{\rm eff}=0.067 m_e$ and a dielectric constant of $\varepsilon_r=12.9$ (for GaAs) one finds $\chi_1 \equiv \chi_{k_F,1}=34.7$cm$^2$/s and $\chi_3 \equiv \chi_{k_F,3}=43.2$cm$^2$/s at $T=0$. 

The two other terms in (\ref{eq:SpinDiffusionEquation}) read in the limit $\alpha=\beta$
\begin{align}
{\bf H}_1 = {}&  2 \tilde D \big\{ \boldsymbol B_1 \tau \times ( \partial_y^2 {\bf S}) + q_0^2 (\boldsymbol B_1 \tau \cdot \hat e_x )({\bf S}\times \hat e_x)\\
& \qquad + q_0 \partial_y [ \boldsymbol B_1 \tau \times({\bf S}\times \hat e_x)]  + q_0 \hat e_x \times(\partial_y {\bf S}\times\boldsymbol B_1 \tau) \nonumber \\
& \qquad + 2 q_0 (\boldsymbol B_1 \tau \cdot\partial_y{\bf S})(\boldsymbol B_1 \tau \times \hat e_x) \nonumber \\
& \qquad + 2 \partial_y[\boldsymbol B_1 \tau(\boldsymbol B_1 \tau \cdot\partial_y{\bf S})] \big\},\nonumber\\
{\bf H}_3 = {} & - 2\tilde\gamma_{\rm cd} (B_{3x} \tau,B_{3y} \tau,0)\times{\bf S}.
\end{align}

Due to the presence of the terms ${\bf H}_{1,3}$ the spin diffusion equation~(\ref{eq:SpinDiffusionEquation}) becomes a nonlinear partial differential equation, because the Hartree-Fock fields ${\bf B}_{1,3}$ depend on the solution for the spin density ${\bf S}$ (see Eq.~(\ref{eq:HartreeFockField})). The exact solution can thus only be found numerically. Note, that for the derivation of the spin diffusion equation only $b(k_F)\tau \ll 1$ was required but no restriction applies with respect to the magnitude of the Hartree-Fock fields $B_{1,3}$, since those fields have winding number zero.

In our previous treatment~\cite{Matthias} of the PSH state in the presence of cubic Dresselhaus spin-orbit interaction we have found that the PSH state survives in the presence of cubic Dresselhaus spin-orbit interaction but the amplitudes of the spin polarization of the PSH-state acquire a small ellipticity and a finite lifetime. The PSH eigenmode of (\ref{eq:hatD}) in the absence of the Hartree-Fock field (i.e.\ with $B_{1,3}\to 0$) reads
\begin{eqnarray}
\label{eq:ConstantPSH}
S_x &=& 0 \\
S_y &=& -S_0 \sin (q_0 y) e^{-t/\tau_{\rm PSH}}\nonumber \\
S_z &=& S_0 \left( \sqrt{1+ \frac{\gamma_{\rm cd}^2}{\Gamma^2}} - \frac{\gamma_{\rm cd}}{{\Gamma}}\right) \cos (q_0 y) e^{-t/\tau_{\rm PSH}} \nonumber \\
& \approx & S_0 \left( 1 - \frac{\gamma_{\rm cd}}{{\Gamma}}\right) \cos (q_0 y) e^{-t/\tau_{\rm PSH}},
\nonumber
\end{eqnarray}
where the lifetime $\tau_{\rm PSH}$ is given by
\begin{equation}
\label{eq:LifetimeConstantPSH}
\frac{1}{\tau_{\rm PSH}}=\frac{1}{2}\Gamma+ \frac{3}{2}\gamma_{\rm cd} - \frac{1}{2}\sqrt{\Gamma^2 + \gamma_{\rm cd}^2} 
\approx \frac{3}{2} \gamma_{\rm cd}.
\end{equation}
Here, we have denoted the D'yakonov-Perel' relaxation rate by $\Gamma=4q_0^2D$ and the approximate expressions for $S_z$ and $\tau_{\rm PSH}$ are valid to lowest order in $\gamma_{\rm cd}/\Gamma \propto [b_{cD}(k_F)/ b_R(k_F)]^2$,  which is basically the square of the ratio of the cubic Dresselhaus spin orbit interaction over the linear spin orbit interactions.

The purpose of the present paper is to investigate how the pattern and the lifetime $\tau_{\rm PSH}$ of the PSH state are modified in the presence of the Hartree-Fock field arising from first-order electron-electron interactions. In Sec.~\ref{sec:PSHlifetime} we present our results concerning the lifetime, based on numerical solution of the diffusion equation (\ref{eq:SpinDiffusionEquation}). Then, in Sec.~\ref{sec:AnalyticalSolution}, we present a more analytical analysis of Eq.~(\ref{eq:SpinDiffusionEquation}). We will simplify the problem and investigate the diffusion equation assuming (i) the limit of small cubic Dresselhaus spin-orbit interaction, $\gamma_{\rm cd}/\Gamma \ll 1$, and (ii) a {\it time-independent} helical magnetic field instead of the self-consistent Hartree-Fock field. These assumptions allow for an analytical approach, making use of the smallness of the parameter $\gamma_{\rm cd}/\Gamma$. Although corresponding to a qualitatively different situation, this approach provides some insight in the general structure of the solutions of Eq.~(\ref{eq:SpinDiffusionEquation}). Finally, in Sec.~\ref{sec:PSHpattern} we present full numerical solutions of Eq.~(\ref{eq:SpinDiffusionEquation}) now focusing on the spatial pattern of the PSH for different initial spin polarizations and for small and large cubic Dresselhaus spin-orbit interaction.

\section{Modifications of the PSH state}
\label{sec:Results}

\subsection{Lifetime enhancement of the PSH state}
\label{sec:PSHlifetime}

\begin{figure}[t]
\includegraphics[height=6cm]{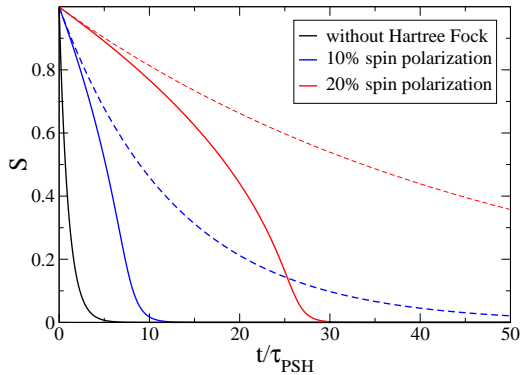}
\caption{\label{fig:Lifetime}
Time dependence of the $S_z$-amplitude of the PSH-state for different initial spin polarizations (solid lines). For comparison the time evolution without Hartree-Fock field is shown in black. Dashed lines indicate, how the time dependence of the $S_z$-amplitude of the PSH-state would look like, if the Hartree-Fock field remained constant at its initial value.}
\end{figure}

Our main finding is, similar to previous work,~\cite{WengWu,Stich} that the PSH lifetime is enhanced considerably due to the Hartree-Fock field arising from first-order electron-electron interactions. The intuitive explanation is simply that the Hartree-Fock field is parallel to the spin polarization of the helix state and thus strengthens the spin polarization. Since, experimentally this can e.g.\ be realized by increasing the initial spin polarization, i.e., the amplitude of the PSH state, we show in Fig.~\ref{fig:Lifetime} the time evolution of the $S_z$ spin polarization for initial spin polarizations of the PSH state with 10\% (blue) and 20\% (red) using the parameters of the quantum well of Ref.~\onlinecite{koralek}: a Fermi temperature of $T_F=400K$, an effective electronic mass of $m_{\rm eff}=0.067 m_e$, linear Rashba and Dresselhaus spin orbit couplings of $\alpha=\beta=0.0013$ and a cubic Dresselhaus spin-orbit interaction of $\gamma v_F=5.0 {\rm eV \AA^3}$ as well as a momentum relaxation time of $\tau=1 {\rm ps}$.
In the simulations we always assume that at time $t=0$ the electronic spin polarization corresponds to the PSH state of Eq.~(\ref{eq:ConstantPSH}). We then solve for the time-dependent polarization ${\bf S}(t)$ with the Hartree-Fock field switched on at time $t=0$.
For comparison the time evolution of the $S_z$ spin polarization of the PSH state {\it without} Hartree-Fock field is also shown (in black). Obviously, for realistic quantum well parameters the lifetime of the PSH state can easily be enhanced by factors of 10 to 20 when the initial spin polarization is in the percentage range.

The main cause for the lifetime enhancement of the PSH-state can be traced back to the reduction of the effective cubic Dresselhaus scattering rate $\tilde \gamma_{\rm cd}$ in Eq.~(\ref{eq:tildeGamma}) in the presence of a Hartree-Fock field. Since it is the SU(2)-symmetry breaking cubic Dresselhaus spin-orbit interaction, that renders the lifetime of the PSH state finite in the first place, a reduction of the effective cubic Dresselhaus scattering rate considerably enhances the lifetime of the PSH state. Because the Hartree-Fock field is proportional to the spin polarization itself, its magnitude and therefore its ability to suppress the effective cubic Dresselhaus scattering rate $\tilde \gamma_{\rm cd}$ decreases with decreasing spin polarization. As a consequence the spin polarization (solid lines) in Fig.~\ref{fig:Lifetime} decays faster than the exponential decay (dashed lines) expected for a constant field.

\subsection{Analytical approach for $\gamma_{\rm cd}/\Gamma \ll 1$}
\label{sec:AnalyticalSolution}

Whereas the full nonlinear solution for the spin polarization in the presence of a self-consistent Hartree-Fock field can only be obtained numerically, we can simplify the problem and work with a static magnetic field instead. We take as an ansatz for the PSH state
\begin{equation}
{\bf S}_{\rm PSH}=S_0 
\begin{pmatrix} 
0\\
-\sin q_0 y \\
(1-x) \cos q_0 y 
\end{pmatrix},
\label{eq:InitialPSHstate}
\end{equation}
where $x$ parametrizes the ellipticity of the spin polarization due to the presence of cubic Dresselhaus spin-orbit interaction (see Eq.~(\ref{eq:ConstantPSH})). We assume the fields ${\bf B}_{1,3}$ to be static and proportional to this polarization profile,
\begin{equation}
{\bf B}_{1,3} = B_{1,3}\frac{{\bf S}_{\rm PSH}}{|{\bf S}_{\rm PSH}|}.
\label{eq:bpsh}
\end{equation}
If we plug the ansatz (\ref{eq:InitialPSHstate}) and the fields (\ref{eq:bpsh}) into the diffusion equation (\ref{eq:SpinDiffusionEquation}), we find that with
\begin{equation}
x= \frac{\tilde \gamma_{\rm cd}}{4q_0^2\tilde D} \frac{1}{1+\frac{1}{2}(B_1 \tau)^2}
\equiv \frac{\tilde \gamma_{\rm cd}}{\tilde \Gamma} \frac{1}{1+\frac{1}{2}(B_1 \tau)^2},
\end{equation}
the instanteneous time-evolution of the PSH state reads to lowest order in $\gamma_{\rm cd}/\Gamma$
\begin{equation}
\label{eq:PSHinitialEvolution}
\partial_t {\bf S}_{\rm PSH}= -\frac{{\bf S}_{\rm PSH}}{\tilde \tau_{\rm PSH}} + \frac{{\bf S}_{2x}}{\tau_{2x}} + \frac{{\bf S}_{3yz}}{\tau_{3yz}},
\end{equation}
where
\begin{align}
\frac{1}{\tilde \tau_{\rm PSH}}= {} & \frac{3}{2} \tilde \gamma_{\rm cd}= \frac{3}{2} \frac{\gamma_{\rm cd}}{1+(2B_3\tau)^2},
\label{eq:InitialLifetime} \\
\frac{{\bf S}_{2x}}{\tau_{2x}} ={} & \frac{S_0}{3 \tilde \tau_{\rm PSH}} \!\! \left( \!\! B_3 \tau - \frac{B_1 \tau}{1+\frac{1}{2}(B_1 \tau)^2} \! \right)\!\!
\begin{pmatrix}
\sin (2 q_0 y) \\
0\\
0
\end{pmatrix},
\label{eq:S2x} \\
\frac{{\bf S}_{3yz}}{\tau_{3yz}} = {} & \frac{S_0}{3 \tilde \tau_{\rm PSH}}\frac{(B_1 \tau)^2}{1+\frac{1}{2}(B_1 \tau)^2} 
\begin{pmatrix}
0\\
-\sin (3 q_0 y)\\
\cos (3 q_0 y)
\end{pmatrix}.
\label{eq:S3yz}
\end{align}
This equation thus describes two effects: (i) The state ${\bf S}_{\rm PSH}$ taken as ansatz tends to relax towards ${\bf S}=0$ with an instantaneous relaxation time $\tilde\tau_{\rm PSH}$, which is enhanced in comparison to the bare $\tau_{\rm PSH}$ due to the reduction of the cubic Dresselhaus scattering rate in the presence of a helical magnetic field. (ii) The polarization starts to develop higher harmonics (with wave vectors $2q_0$ and $3q_0$), including an $x$-component which was not present before (cf.~Eq.~(\ref{eq:ConstantPSH})).

As we will show below, the magnitudes of the higher harmonics are smaller by factors $\gamma_{\rm cd}/\Gamma$ (see Eq.~(\ref{eq:sxSolution})). This means that in the limit of $\gamma_{\rm cd}/\Gamma \ll 1$ we can neglect the contributions of ${\bf S}_{2x}$ and ${\bf S}_{3yz}$ in (\ref{eq:PSHinitialEvolution}) to first approximation, and Eq.~(\ref{eq:PSHinitialEvolution}) becomes valid at all times (at least for constant magnetic fields $B_{1,3}$). In this case one can solve the equation easily to find the simple exponential decay
\begin{equation}
{\bf S}_{\rm PSH}(t)= {\bf S}_{\rm PSH} e^{-t/\tilde \tau_{\rm PSH}}.
\end{equation}
This exponential solution for constant fields is indicated by dashed lines in Fig.~\ref{fig:Lifetime}. For the actual time-dependent Hartree-Fock field (solid lines in Fig.~\ref{fig:Lifetime}) the lifetime is of course shorter than $\tilde \tau_{\rm PSH}$ of Eq.~(\ref{eq:InitialLifetime}): as soon as the spin polarization of the PSH state starts to decay, the fields ${\bf B}_{1,3}$ decrease proportionally, which results in an increase of the decay rate $1/\tilde \tau_{\rm PSH}$ of the PSH state.

Let us now turn to the generation of higher harmonics in (\ref{eq:PSHinitialEvolution}), which is due to the nonlinearity of the spin diffusion equation. In particular a second harmonic $\sin (2 q_0 y)$-term appears in the $S_x$ component (see Eq.~(\ref{eq:S2x})) and a third harmonic $\sin(3 q_0 y)$-, $\cos (3 q_0 z)$-term appears in the $S_y$- and $S_z$-components (see Eq.~(\ref{eq:S3yz})). Of course, this is not the end of the story, since, in order to obtain the full solution, the second and third harmonics would have to be included into the initial expression leading to a successive generation of all higher harmonics.

The most surprising and at first counterintuitive finding is the generation of a finite $S_x$-component of the spin helix, i.e., the observation that a Hartree-Fock field parallel to the local spin polarization rotates the spin polarization of the PSH out of the $yz$-plane by creating an additional out-of-plane $S_x$-component. In order to estimate the magnitude of this new second harmonic $S_x$-component of the PSH-state, we modify the ansatz of the initial PSH state of Eq.~(\ref{eq:InitialPSHstate}) to include the second harmonic $S_x$-component as
\begin{equation}
{\bf S}_{{\rm PSH}2x}=S_0 
\begin{pmatrix} 
s_x \sin 2 q_0 y\\
-\sin q_0 y \\
(1-x) \cos q_0 y 
\end{pmatrix}.
\label{eq:InitialPSHstateS2x}
\end{equation}
Plugging this ansatz into the spin diffusion equation~(\ref{eq:SpinDiffusionEquation}) one finds for the $x$-component of ${\bf S}_{{\rm PSH}2x}$ to leading order in $\gamma_{\rm cd}/\Gamma$
\begin{equation}
\partial_t (S_0 s_x) = S_0\frac{\tilde \gamma_{\rm cd}}{2} \Bigl(  B_3 \tau - \frac{B_1 \tau}{1+\frac{1}{2}(B_1 \tau)^2} \Bigr)- \frac{\tilde \Gamma}{4} S_0 s_x.
\label{eq:dts0sx}
\end{equation}
We see that the second harmonic has a fast relaxation rate $\tilde \Gamma$ compared to the time scale $1/\tilde\gamma_{\rm cd}$ of the other terms. Let us solve (\ref{eq:dts0sx}) for $s_x(0)=0$, i.e., starting from the PSH ansatz of Eq.~(\ref{eq:InitialPSHstate}). Assuming that the spin polarization of the PSH state $S_0$ remains constant in time we find
\begin{equation}
s_x(t)=\frac{\tilde \gamma_{\rm cd}}{2 \tilde \Gamma}  \Bigl(  B_3 \tau - \frac{B_1 \tau}{1+\frac{1}{2}(B_1 \tau)^2} \Bigr) \left (1-e^{-\frac{1}{4}\tilde \Gamma t} \right).
\label{eq:sxSolution}
\end{equation}
From this simple consideration we see that the fast relaxation rate of $s_x$ has two effects: (i) It makes the equilibrium amplitude of the $S_x$-component very small for small cubic Dresselhaus spin-orbit interaction, since it is proportional to $\tilde \gamma_{\rm cd}/\tilde \Gamma$. (ii) It causes the $S_x$-component to rise from zero to its maximum value within a short switch on period given by $1/\tilde \Gamma \ll \tilde \tau_{\rm PSH}$.

Can we understand the physical origin of this second harmonic $S_x$-component? Normally, one would expect that a parallel magnetic field simply strengthens the parallel spin orientation. In the presence of spin-orbit fields, however, the spin density consists of parts with different winding numbers, where only winding number zero contributes to the local spin polarization while the other winding numbers (one and three) average out to zero. Since the anisotropic components of the spin density (winding numbers one and three) are not necessarily parallel to the local spin density and thus not necessarily parallel to the Hartree-Fock field, they can modify the pattern of the PSH state by rotations around the local Hartree-Fock field.

\begin{figure}[t]
\includegraphics[height=5cm]{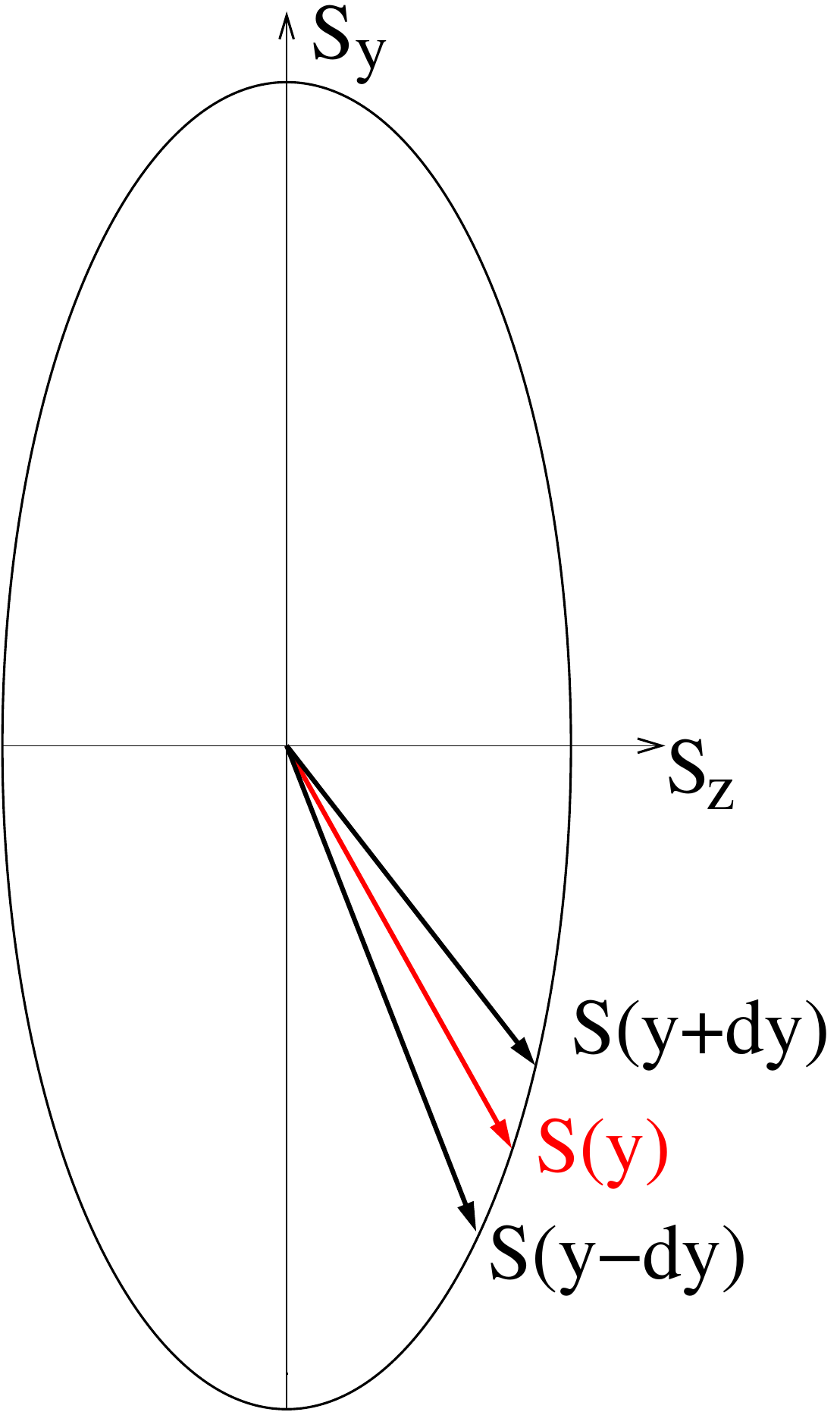}
\caption{\label{fig:Ellipse}
For equal amplitudes $S(y+dy)=S(y-dy)$ the contribution of the winding number one spins to the $S_x$-component diffusing from $y+dy$ and from $y-dy$ to the point $y$ would cancel, because they rotate around the same local Hartree Fock field $B(y)$ but correspond to $\pm k$-states and thus are oriented oppositely. Due to the elliptical profile of the PSH amplitude the amplitudes $S(y+dy)$ and $S(y-dy)$ differ and thus a finite $S_x$-contribution emerges.}
\end{figure}

The expression for the second harmonic $S_x$-component in the presence of a constant Hartree-Fock field (see Eq.~(\ref{eq:sxSolution})) suggests that both winding number one and winding number three parts of the spin density contribute to the formation of the $S_x$-component. The winding number one part contributes only via the diffusion term. At one fixed position $y$ electronic spins arrive via diffusion from $y +\delta y$ with amplitude ${\bf S}(y + \delta y)$ and from $y-\delta y$ with amplitude ${\bf S}(y - \delta y)$. While diffusing to position $y$ both spins arriving from $y\pm \delta y$ rotate around the local Hartree-Fock field, which is parallel to  ${\bf S}(y)$ and thus acquire a finite $S_x$-component, since ${\bf S}$ is polarized in the $yz$-plane (See Fig.~\ref{fig:Ellipse}). Since the angle between  ${\bf S}(y + \delta y)$ and ${\bf S}(y)$ is equal to the angle between ${\bf S}(y)$ and ${\bf S}(y-\delta y)$ the resulting $S_x$-components are of opposite sign an would cancel for equal magnitudes $S(y+\delta y)=S(y-\delta y)$, i.e., for a PSH state with equal $S_y$ and $S_z$ amplitudes.
However, in the presence of cubic Dresselhaus spin orbit interaction the initial PSH profile is elliptical with the $S_y$ and $S_z$ amplitudes differing by approximately $\gamma_{\rm cd}/\Gamma$. Thus the amplitudes of the electronic spins arriving from $y \pm \delta y$ differ due to the elliptical PSH profile leading to the formation of a finite $S_x$-component (see Fig.~\ref{fig:Ellipse}). Only when the local spin polarization is parallel to one of symmetry axes of the ellipse, i.e., for $S_y=0$ or $S_z=0$ the amplitude of the electronic spins arriving from $y \pm \delta y$ are equal and thus $S_x=0$ at these points. Since this situation occurs for $q_0 y= 0, \pi/2, \pi, 3\pi/2$ this explains the oscillation of $S_x$ with double wave vector $2q_0$.

The winding number three part of the spin distribution function contributes to the formation of an $S_x$-component via combined rotations around the local Hartree-Fock field and the component of the cubic Dresselhaus spin orbit field parallel to $\hat e_y$. Again a finite contribution to $S_x$ arises only when the local spin polarization is neither parallel nor perpendicular to $\hat e_y$ (the direction of the relevant component of the cubic Dresselhaus spin orbit field) thus leading to an oscillation of the winding number three contribution to $S_x$ with double wave vector $2q_0$ similar to the winding number one case.~\cite{WindingNumber3Explanation}

\subsection{Modification of PSH pattern}
\label{sec:PSHpattern}

\begin{figure}[t]
\begin{minipage}{.44\columnwidth}
\hspace*{-.7cm}
\includegraphics[height=3.3cm,clip=true]{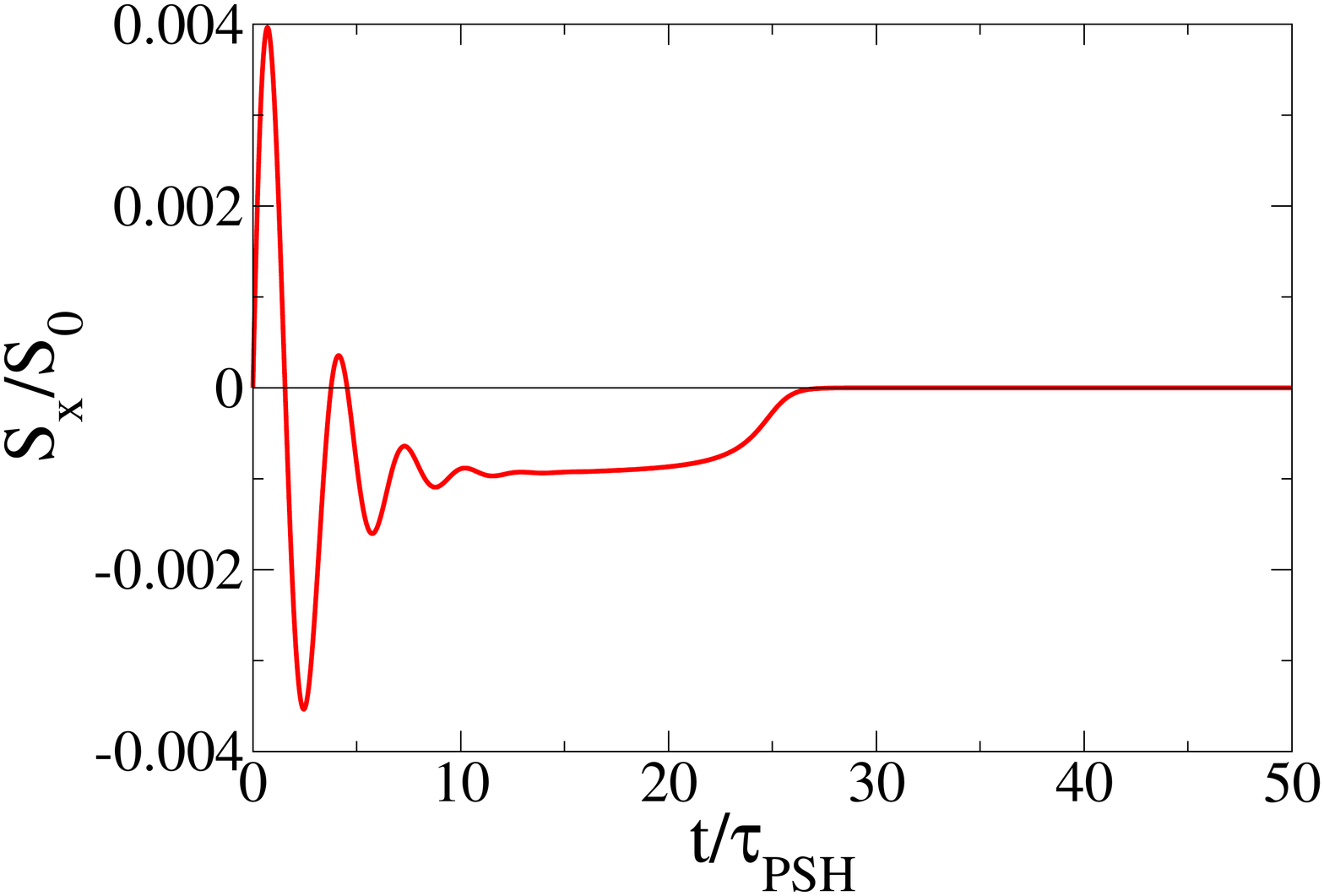}
\end{minipage}
\begin{minipage}{.44\columnwidth}
\includegraphics[height=3.3cm]{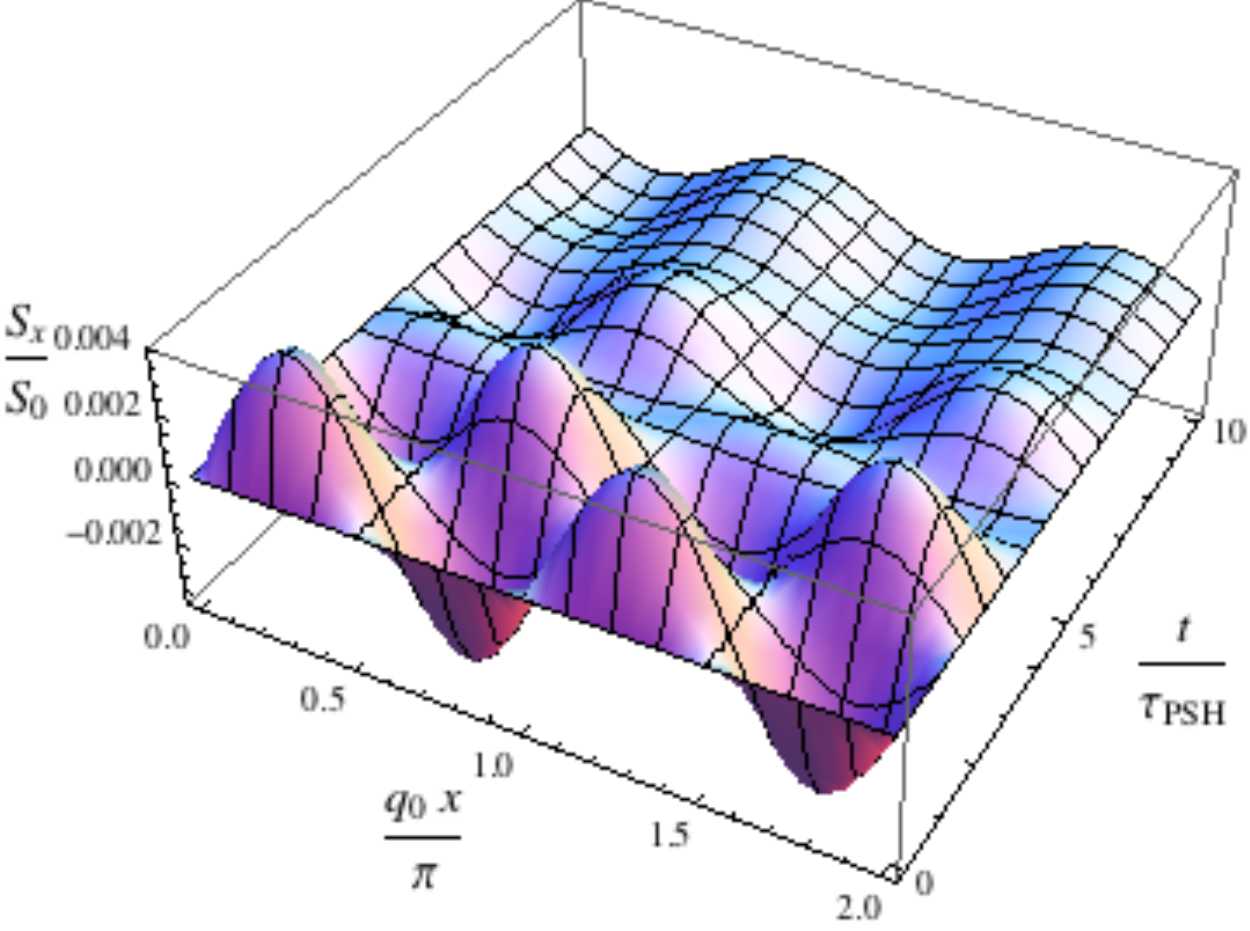}
\end{minipage}
\caption{\label{fig:GcdalphaF02}
Time dependence of the $S_x$-amplitude, which arises solely due to the Hartree Fock term, at first maximum $q_0 y = \pi/2$ (left) and spatial and temporal evolution of the $S_x$-amplitude of the PSH state (right) for an initial spin polarization of 20\% and with linear spin-orbit interactions reduced by a factor of 0.2.  ($\gamma_{\rm cd}/\Gamma =0.067$ and $\tau_{\rm PSH}=412 {\rm ps}$).}
\end{figure}

Since the magnitude of the $S_x$-component as well as the other higher harmonics are of higher order in $\gamma_{\rm cd}/\Gamma$, they will become significant for systems with large cubic Dresselhaus spin orbit interaction. Therefore it is interesting to investigate how the spatial and temporal evolution of the PSH state is modified when cubic Dresselhaus spin orbit interaction is large $\gamma_{\rm cd} \approx \Gamma$. Since the magnitude of the cubic Dresselhaus spin orbit interaction is given by the crystal symmetry of GaAs, whereas the magnitudes of the linear Rashba and Dresselhaus spin orbit interactions $\alpha$ and $\beta$ can be varied by changing the doping asymmetry and the width of the quantum well, we could access the $\gamma_{\rm cd} \approx \Gamma$-regime by reducing the magnitude of the linear spin orbit interactions while keeping the magnitude of the cubic Dresselhaus spin-orbit interaction fixed. Since the lifetime of the PSH depends mainly on the magnitude of cubic Dresselhaus spin-orbit interaction (see Eq.~(\ref{eq:ConstantPSH}), where one finds $\tau_{\rm PSH}^{-1} \approx \frac{3}{2} \gamma_{\rm cd}$ for $\gamma_{\rm cd} \ll \Gamma$ and $\tau_{\rm PSH}^{-1} \approx \gamma_{\rm cd}$ for $\gamma_{\rm cd} \gg \Gamma$), this does not significantly affect the absolute lifetime of the PSH state. Alternatively, the regime of large cubic Dresselhaus spin-orbit interaction could also be reached by using an (In,Ga)As quantum well instead, or by increasing the Fermi energy.

In Fig.~\ref{fig:GcdalphaF02} we show the temporal and spatial evolution of the $S_x$-component of the PSH state, which arises solely due to the Hartree-Fock term, for only slightly enhanced cubic Dresselhaus spin orbit interaction. This situation has been reached by a reduction of the linear spin orbit interactions by a factor of $0.2$, which results in a ratio of $\gamma_{\rm cd}/\Gamma =0.067$. As expected from Eq.~(\ref{eq:sxSolution}) the magnitude of the $S_x$-component is quite small and its spatial variation is characterized by a $\sin (2 q_0 y)$-dependence. The initial switch on process now exhibits an oscillatory behavior and, contrary to the $S_y$- and $S_z$-components, the magnitude of the $S_x$-component saturates before it drops to zero quite abruptly.

\begin{figure}[t]
\begin{minipage}{.44\columnwidth}
\hspace*{-.5cm}
\includegraphics[height=3.3cm]{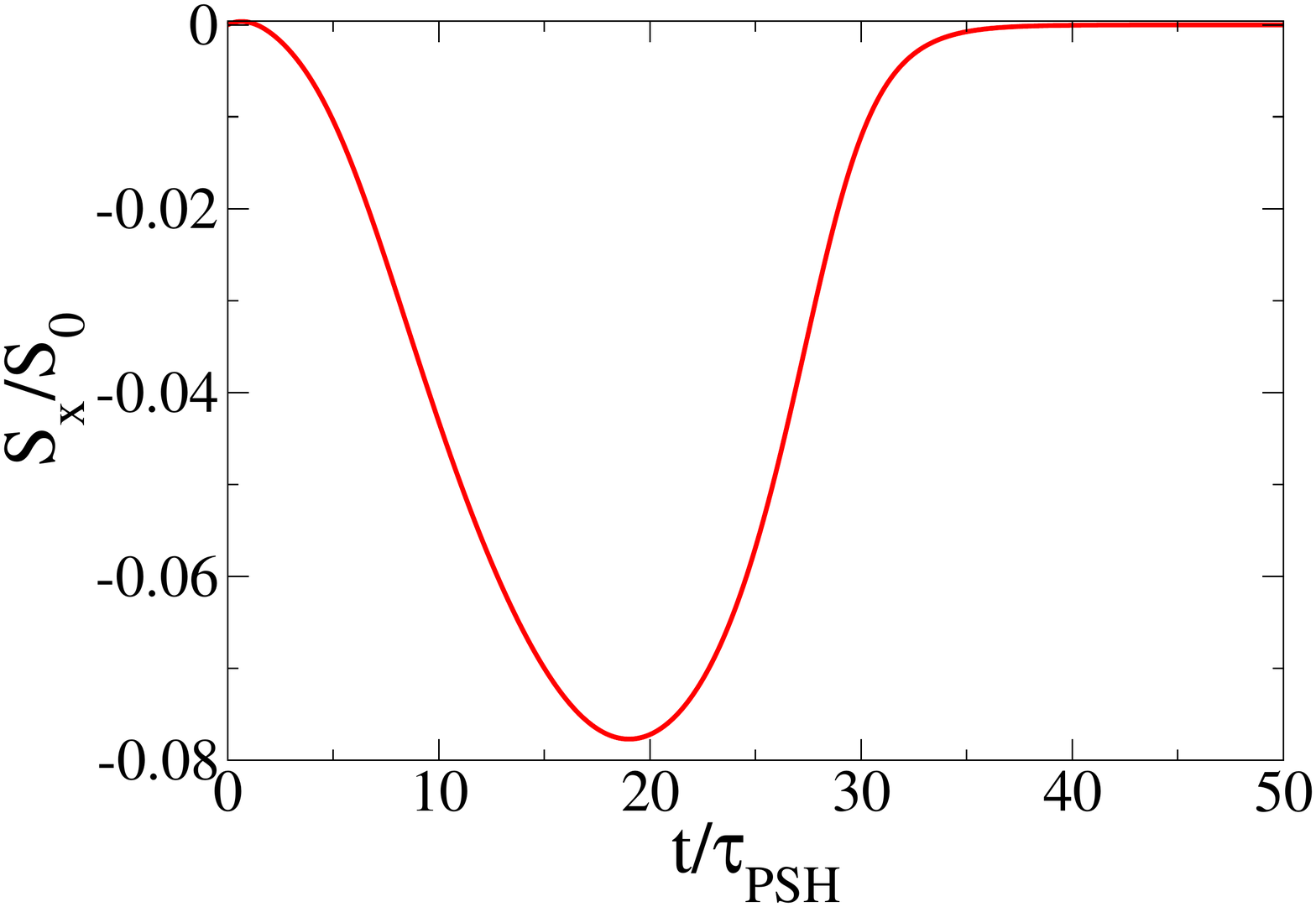}
\end{minipage}
\begin{minipage}{.44\columnwidth}
\includegraphics[height=3.3cm]{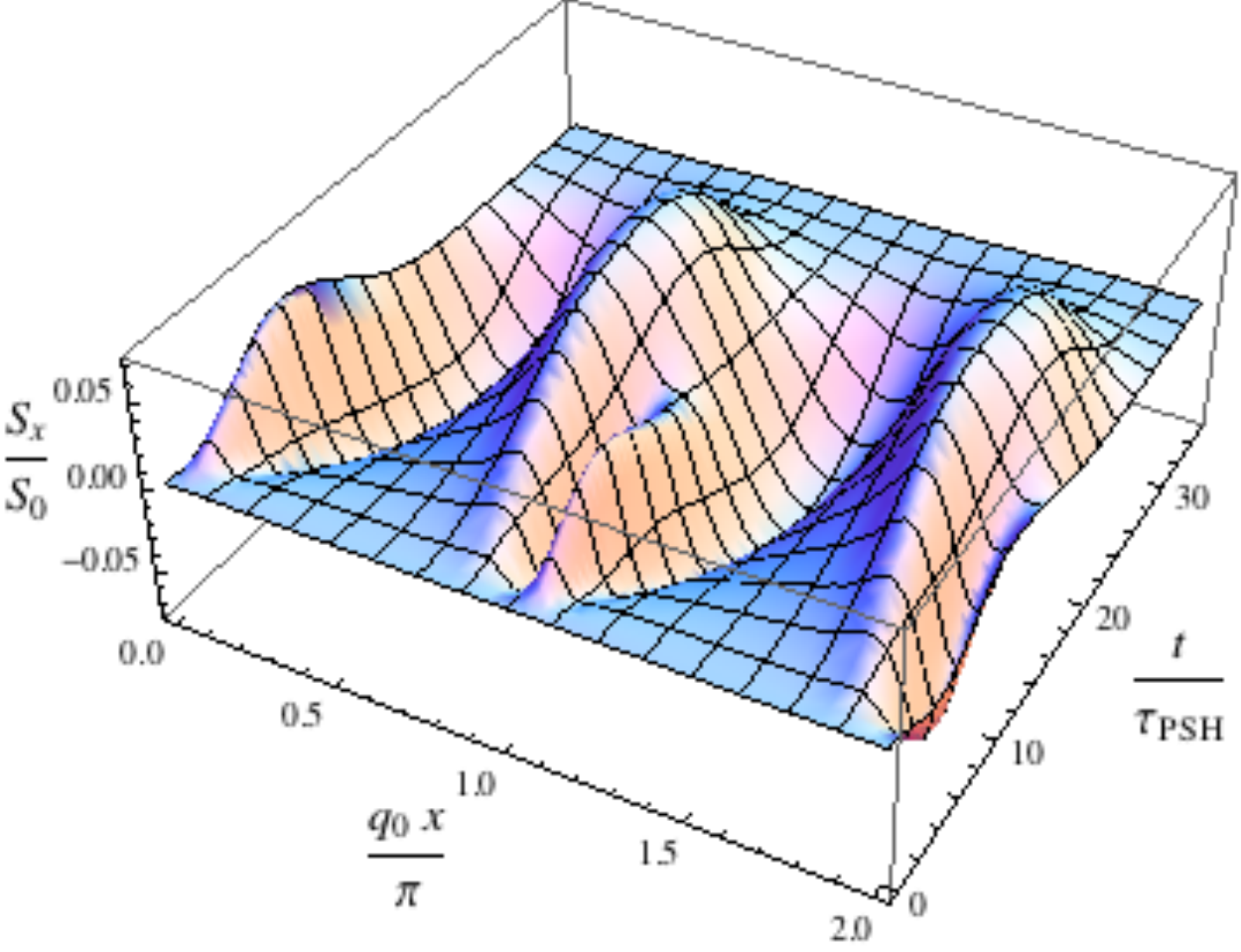}
\end{minipage}
\caption{\label{fig:GcdalphaF002}
Time dependence of the $S_x$-amplitude at first maximum $q_0 y = \pi/2$ (left) and spatial and temporal evolution of the $S_x$-amplitude (right) of the PSH state for an initial spin polarization of 20\% and with linear spin-orbit interactions reduced by a factor of 0.02. ($\gamma_{\rm cd}/\Gamma =6.8$ and $\tau_{\rm PSH}=324 {\rm ps}$).}
\end{figure}

\begin{figure}[t]
\begin{minipage}{.44\columnwidth}
\hspace*{-.5cm}
\includegraphics[height=3.3cm]{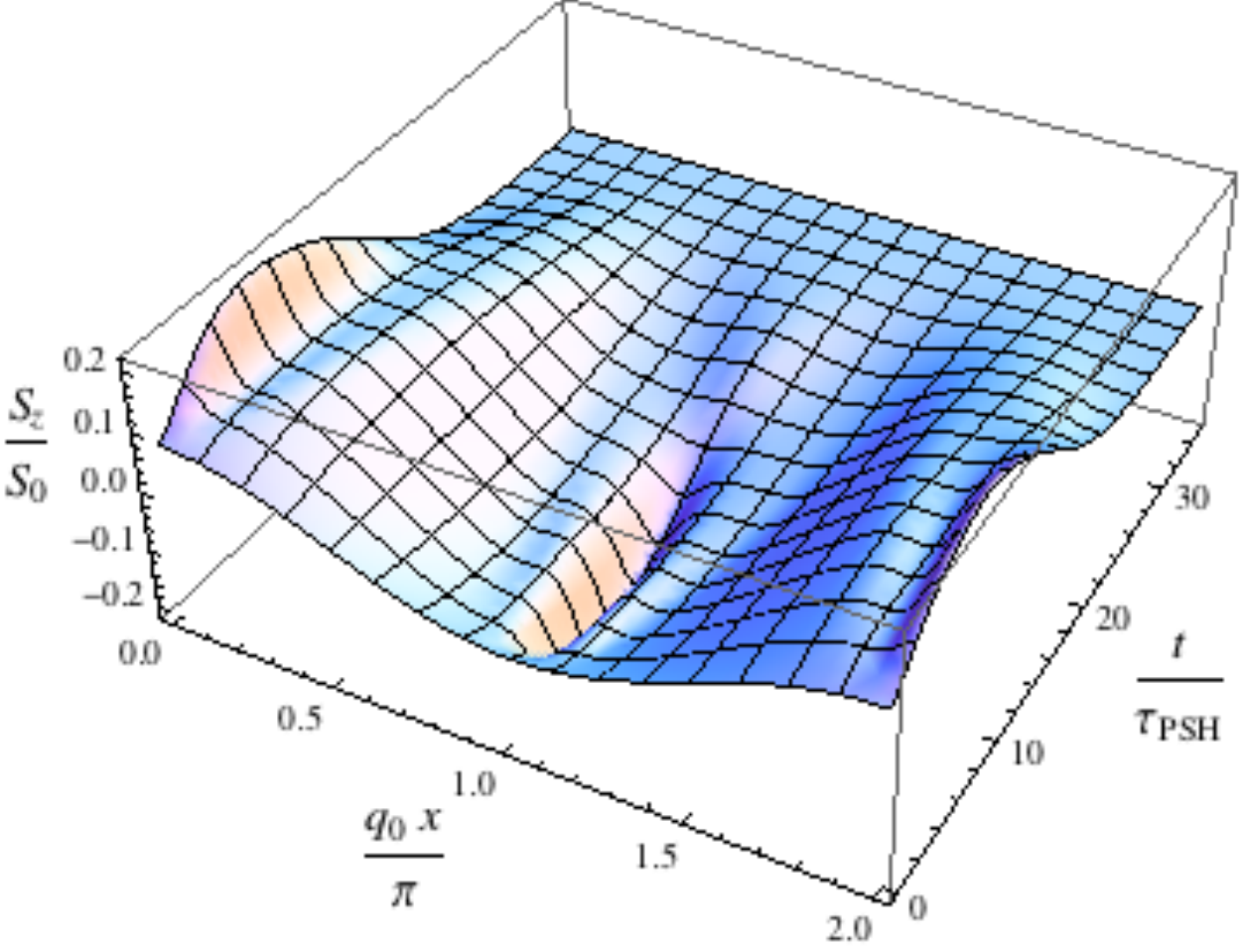}
\end{minipage}
\begin{minipage}{.44\columnwidth}
\includegraphics[height=3.3cm]{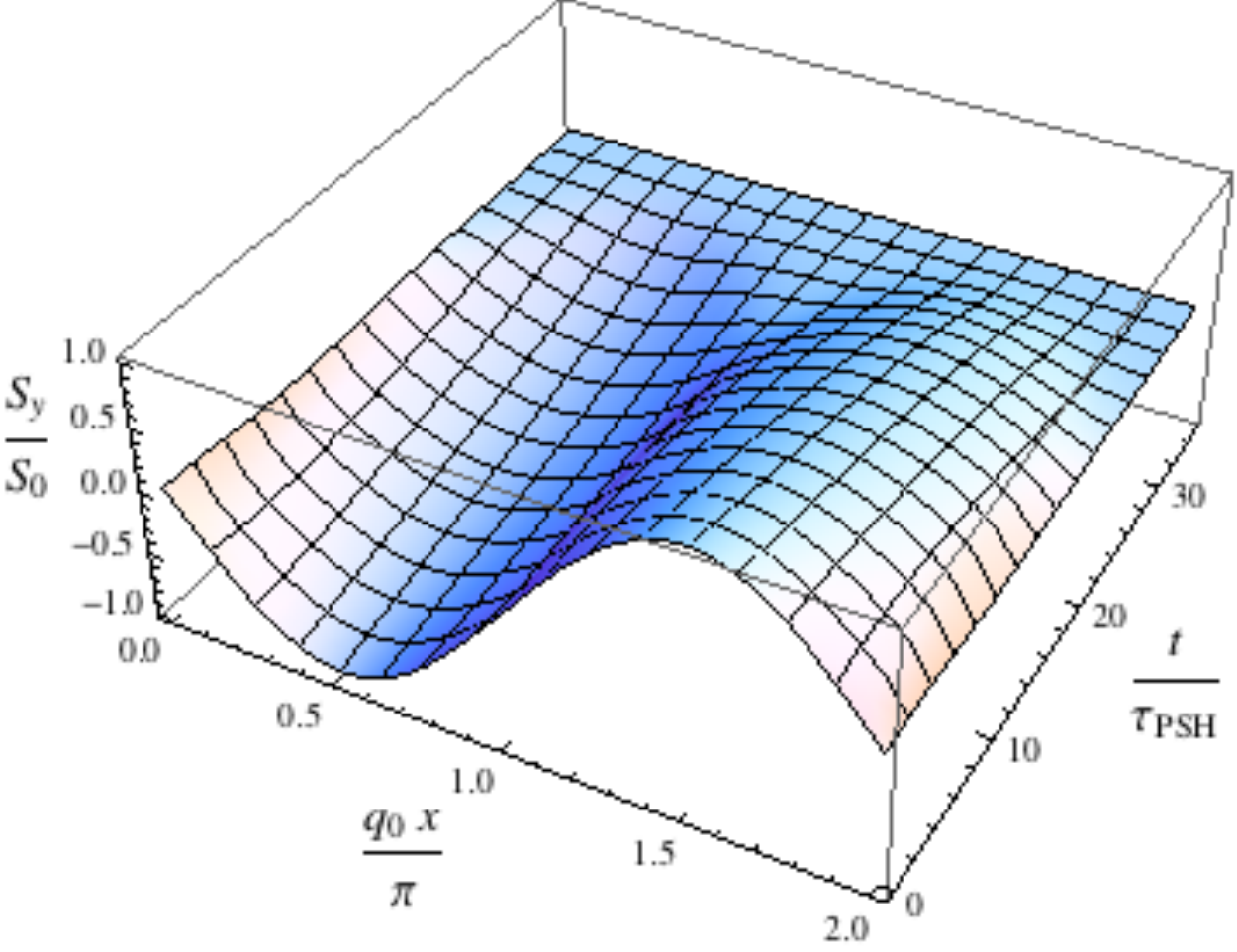}
\end{minipage}
\caption{\label{fig:GcdalphaF002SySz}
Spatial and temporal evolution of the $S_z$-component (left) and the $S_y$-component (right) of the PSH state for the same parameters as used in Fig.~\ref{fig:GcdalphaF002}.}
\end{figure}

As an example for large cubic Dresselhaus spin orbit interaction we show in Fig.~\ref{fig:GcdalphaF002} the spatial and temporal evolution of the PSH state for linear spin-orbit interactions reduced by a factor of $0.02$, which results in a ratio of $\gamma_{\rm cd}/\Gamma=6.8$. The magnitude of the $S_x$-component is now quite sizable, reaching almost 8\% of the PSH amplitude at intermediate times. Obviously, the spatial variation is no longer characterized by a simple $\sin (2 q_0 y)$-dependence but even higher harmonics start to play a prominent role. For comparison the spatial and temporal evolution of the $S_y$- and $S_z$-components are shown in Fig.~\ref{fig:GcdalphaF002SySz}. Due to the large ratio of $\gamma_{\rm cd}/\Gamma=6.8$ also the amplitude of the $S_z$-component is strongly reduced and the presence of higher harmonics is clearly visisble.

\section{Conclusions}
Based on a spin-coherent Boltzmann type approach we have derived a semiclassical spin diffusion equation for a two-dimensional electron gas with Rashba- and Dresselhaus spin-orbit interactions including the Hartree-Fock field, which arises in first order in the  electron-electron interactions. In agreement with previous treatments~\cite{Stich,WengWu,Takahashi} we find that the Hartree-Fock term considerably enhances the lifetime of the persistent spin helix state, when the initial spin polarization is in the percentage range. Within our approach we are able to assign the origin of the lifetime enhancement to the suppression of the effective cubic Dresselhaus scattering rate by the Hartree Fock field.

Surprisingly, not only the lifetime but also the pattern of the spin polarization is modified in the presence of a Hartree Fock field. Due to the nonlinearity of the problem higher harmonics in the spin polarization are generated. The second harmonic corresponds to a spin polarization in $S_x$-direction and thus rotates the spin polarization of the persistent spin helix state out of the $(S_y,S_z)$-plane. These modifications of the spin pattern of the persistent spin helix become significant when the cubic Dresselhaus spin-orbit interaction is of similar magnitude as the linear spin-orbit interactions. Since the magnitude of the cubic Dresselhaus spin-orbit interaction is fixed by crystal symmetry, we have suggested that this regime could be experimentally reached by reducing the magnitude of the linear spin-orbit interactions via appropriate design of the quantum wells.

\begin{acknowledgments}

We would like to thank P.W. Brouwer for helpful discussions. This work was supported by SPP 1285 of the DFG and by the
Alexander von Humboldt Foundation.

\end{acknowledgments}

\begin{appendix}

\section{Hartree Fock term}
\label{app:HartreeFock}

For a general nonequilibrium problem with the Hamiltonian $H=H_0+V$ containing an exactly solvable single-particle part $H_0$ and an interaction $V$, Zubarev's Nonequilibrium statistical operator formalism\cite{zubarev} allows to derive a closed set of equations that describe  the irreversible temporal evolution of the density matrix $\langle \psi^\dagger_{l'}\psi_{l} \rangle^t=\textnormal{Tr}[\rho(t)\,\psi^\dagger_{l'}\psi_{l}]$. Here $l,l'$ are possibly composed indices, e.g., for momentum and spin. 
For details of the derivation, see Ref.~\onlinecite{zubarev}. Since for zero temperature the 2nd order electron-electron interaction vanishes we consider the kinetic equation only up to first order in $V$, yielding:
\begin{equation}
\label{kineqmarkov}
\partial_t \langle \psi^\dagger_{l'}\psi_{l} \rangle^t-i\,\langle[H_0,\psi^\dagger_{l'}\psi_{l}]\rangle^t =  {J}^{\mathrm{(1)}}_{l l'}(t),
\end{equation}
with the first order mean field term:
\begin{equation}
{J}^{\mathrm{(1)}}_{l l'}(t) = i\langle[V,\psi^\dagger_{l'}\psi_{l}]\rangle_\textnormal{rel}^t.
\label{eq:meanfield}
\end{equation}

We evaluate the mean field term (\ref{eq:meanfield}) for the electron-electron interactions of Eq.~(\ref{eq:Hee}). Evaluation of the commutator in  Eq.~(\ref{eq:meanfield}) yields in this notation
\begin{eqnarray}
{J}^{\rm ee(1)}_{11'} \!\!\!&=&\!\! i \,\langle [H_\textnormal{e-e},\psi^\dagger_{1'}\psi_{1}]\rangle \\
  &=& \!\! \frac{1}{i V} \!\! \sum_{2' 3' 2 3} \!\!\! V_{2'3' 23} \! \left(\! \delta_{1 3'}\! \langle \psi^\dagger_{1'}\psi^\dagger_{2'} \psi_{2}\psi_{3}\rangle \!-\! \delta_{1' 3} \langle \psi^\dagger_{3'}\psi^\dagger_{2'} \psi_{2} \psi_{1}\rangle \! \right). \nonumber 
\end{eqnarray}
Upon Wick decomposition of the averages 
\begin{eqnarray}
\label{wickapplied}
\langle \psi^\dagger_{1}\psi^\dagger_{2} \psi_{3} \psi_{4}\rangle &=& - \langle \psi^\dagger_{1}\,\psi_{3}\rangle \,\langle \psi^\dagger_{2}\,\psi_{4}\rangle+ \langle \psi^\dagger_{1} \psi_{4}\rangle \,\langle \psi^\dagger_{2} \psi_{3}\rangle\nonumber \\
    &=& - \delta_{{\bf k}_1 {{\bf k}_3}} \delta_{{{\bf k}_2} {{\bf k}_4}} f_{s_1 s_3}({{\bf k}_1}) f_{s_2 s_4}({{\bf k}_2})\nonumber  \\
   && + \delta_{{{\bf k}_1}{{\bf k}_4}} \delta_{{{\bf k}_2} {{\bf k}_3}} f_{s_1 s_4}({{\bf k}_1}) f_{s_2 s_3}({{\bf k}_2})
\end{eqnarray}
and by exploiting the Kronecker symbols we obtain  a mean field term that is diagonal in momentum but remains a matrix in spin space
\begin{equation}
{J}^{\rm ee(1)}_{s s'}\!({\bf k}) \!=\! \frac{1}{i V} \! \sum_{{\bf k'}} \! \tilde{v}(|{\bf k'} \!- {\bf k}|) \! \left[ f({\bf k'}) f({\bf k}) - f({\bf k}) f({\bf k'}) \right]_{s s'}.
\end{equation}
By decomposing   the spin space matrices according to
\begin{equation}
     \hat{f}({\bf k}) = n_{\bf k}+{\boldsymbol{\sigma}}\cdot {\bf s}_{\bf k}\,,\qquad
     {\hat{J}}^{\rm ee(1)}_{\bf k}  = {{J}}^{\rm ee(1)}_{\bf k}+{\boldsymbol{\sigma}}\cdot {{\bf J}}^{\rm ee(1)}_{\bf k}\,,
\end{equation}
one finds that the mean field term entering on the right-hand side of the kinetic equation for the  spin density (Eq.~(\ref{eq:ChargeSpin})) becomes:
\begin{equation}
{\bf J}_{\bf k}^{\rm ee(1)} = - 2\, {\bf s}_{\bf k} \times \int \frac{d {\bf q}}{(2 \pi)^2} \tilde v(q) {\bf s}_{\bf k+q}.
\label{eq:1storderEECollisionIntegral}
\end{equation}

\section{Spin diffusion equation}
\label{app:SpinDiffusionEquation}

We follow the approach used in our previous work~\cite{Matthias}  to set up a diffusion equation for the spin density ${\bf s_k}$ valid in the diffusive D'yakonov-Perel' regime, i.e., in the regime $b_F \tau \ll 1$ where due to strong scattering the spin polarization is stabilized. 
In order to solve equation~(\ref{eq:ChargeSpin}) for the spin density we expand the spin density into k-space winding numbers
\begin{equation}
{\bf s_k}=\sum_{l=0,\pm1,\pm 3} {\bf s}_l(k) e^{i l \theta},
\end{equation}
where we keep besides the isotropic component ${\bf s}_0=-\frac{2 \pi}{m} f'(\epsilon_k) {\bf S}$ anisotropic components with winding numbers $\pm 1$ and $\pm 3$. Other winding numbers would be of higher order in ${b(k_F)} \tau$ and therefore need not to be considered in the diffusive regime.

Using this expansion of the spin density, the 1st order electron-electron collision integral (Eq.~(\ref{eq:1storderEECollisionIntegral})) reduces in the diffusive regime to: 
\begin{equation}
{\bf J}_{\bf k}^{\rm ee(1)} = -{\bf S} \times ( \chi_{{\bf k},1} {\bf s}_{{\bf k},\pm 1} + \chi_{{\bf k},3} {\bf s}_{{\bf k},\pm 3}),
\label{eq:HartreeFockPrecession}
\end{equation}
where $\chi_1$ and $\chi_3$ have been defined in Eq.~(\ref{eq:Chi13}).

 Plugging the Hartree-Fock precession term~(\ref{eq:HartreeFockPrecession}) back into the equation for the spin density~(\ref{eq:ChargeSpin}) one finds for the isotropic components of the spin density 
\begin{equation}
\partial_t {\bf s}_0 \!=\! -\frac{v}{2} \partial_x {\bf s}_c - \frac{v}{2} \partial_y {\bf s}_s - {\bf s}_c \times {\bf b}_c - {\bf s}_s \times {\bf b}_s - {\bf s}_{c3} \times {\bf b}_{c3}  - {\bf s}_{s3} \times {\bf b}_{s3} \\ 
\label{eq:s0}
\end{equation}
with
\begin{eqnarray}
{\bf s}_{c} &=& {\bf s}_1+{\bf s}_{-1},\, {\bf s}_{c3}={\bf s}_3+{\bf s}_{-3} \\
{\bf s}_s &=& i({\bf s}_1-{\bf s}_{-1}),\, {\bf s}_{s3}=i({\bf s}_3-{\bf s}_{-3}) \nonumber
\end{eqnarray}
and the spin-orbit fields
\begin{eqnarray}
{\bf b}_c = v_F k (-\alpha + \beta) \hat {\bf e}_y &,&\, {\bf b}_s= v_F k (\alpha + \beta) \hat {\bf e}_x, \\
{\bf b}_{c3} = -\gamma v_F \frac{k^3}{4} \hat {\bf e}_y &,&\, {\bf b}_{s3}=\gamma v_F \frac{k^3}{4} \hat {\bf e}_x. \nonumber
\label{eq:SOI}
\end{eqnarray}

In order to obtain a closed equation for ${\bf s}_0$ one has to determine ${\bf s}_c$, ${\bf s}_s$, ${\bf s}_{c3}$ and ${\bf s}_{s3}$ from the anisotropic components of the Boltzmann equation. In the diffusive regime it is sufficient to find the (quasi)-equilibrium solutions for the anisotropic coefficients, which are obtained by omitting the time derivative of the anisotropic components. The justification for doing so is that, in order to capture the slow precession-diffusion dynamics of the real space density, we can interpret the time derivative as a coarse-grained one, i.e.~$\partial_t \,{\bf s}_0\rightarrow \Delta {\bf s}_0/\Delta t$ with $\Delta t\approx b_F^{-1} \gg \tau$.  Then the fast relaxation of the anisotropic components into the  steady state at the beginning of each time interval $\Delta t$ contributes only in higher order in $b_F\,\tau$ to the average over $\Delta t$. Using these approximations one finds for the winding number one components 
\begin{eqnarray}
\label{eq:ScSs}
{\bf s}_c &=& - \tau_1 v \partial_x {\bf s}_0 + 2 \tau_1 {\bf b}_c \times {\bf s}_0+ 2 \tau_1 {\bf B}_1 \times {\bf s}_c \\
{\bf s}_s &=& - \tau_1 v \partial_y {\bf s}_0 + 2 \tau_1 {\bf b}_s \times {\bf s}_0 + 2 \tau_1 {\bf B}_1 \times {\bf s}_s, \nonumber 
\end{eqnarray}
where ${\bf B}_1=\chi_1 {\bf S}$ is the Hartree-Fock field felt by the winding number one spins and $\tau_1$ is the effective relaxation time for the winding number one part of the spin distribution function. Although in our model $\tau_1=\tau$, this is not true in general, e.g. consideration of 2nd order electron-electron scattering~\cite{Matthias} will add an additional term to $\tau_1$. Therefore we use the more general $\tau_1$ in all our expressions in order to make an extension to including e.g. 2nd order electron-electron interaction straightforward.

Solving Eqs.~(\ref{eq:ScSs}) one finds 
\begin{eqnarray}
\label{eq:SolutionWindingNumber1}
{\bf s}_c &=& - \tilde \tau_1 \left\{ v \partial_x {\bf s}_0 -2 {\bf b}_c \times {\bf s}_0 + 2 \tau_1 v ({\bf B}_1\!\! \times \! \partial_x {\bf s}_0) \right. \\
&& \left. \qquad - 4 \tau_1 {\bf B}_1\!\! \times \! ( {\bf b}_c \! \times \! {\bf s}_0) + 4 \tau_1^2 v {\bf B}_1 ({\bf B}_1 \cdot \partial_x {\bf s}_0) \right\} \nonumber \\
{\bf s}_s &=& - \tilde \tau_1 \left\{ v \partial_y {\bf s}_0 -2 {\bf b}_s \times {\bf s}_0 + 2 \tau_1 v ({\bf B}_1\!\! \times \! \partial_y {\bf s}_0) \right. \nonumber \\
&& \left. \qquad - 4 \tau_1 {\bf B}_1\!\! \times \! ( {\bf b}_s \! \times \! {\bf s}_0)  + 4 \tau_1^2 v {\bf B}_1 ({\bf B}_1 \cdot \partial_y {\bf s}_0) \right\}, \nonumber
\end{eqnarray}
where we have used ${\bf B}_1 || {\bf s}_0$ and
\begin{equation}
\tilde \tau_1 = \frac{\tau_1}{1+(2 B_1 \tau_1)^2}.
\end{equation}

Similarly, the steady state of the winding number three anisotropic components of the spin density is governed by the following equations
\begin{eqnarray}
\label{eq:Sc3Ss3}
\frac{{\bf s}_{c3}}{\tau_3} &=& 2 {\bf b}_{c3} \times {\bf s}_0 + 2 {\bf B}_3 \times {\bf s}_{c3} \\
\frac{{\bf s}_{s3}}{\tau_3} &=& 2 {\bf b}_{s3} \times {\bf s}_0 + 2 {\bf B}_3 \times {\bf s}_{s3}, \nonumber
\end{eqnarray}
where ${\bf B}_3=\chi_3 {\bf S}$ is the Hartree-Fock field felt by the winding number three spins and $\tau_3$ is the effective relaxation time for the winding number three part of the spin distribution function. Although in our model $\tau_3=\tau$ we keep the more general $\tau_3$ in all our expressions based on the same reasoning as detailed for $\tau_1$ below Eq.~(\ref{eq:ScSs}).

Solving the equations for ${\bf s}_{c3}$ and ${\bf s}_{s3}$ one finds
\begin{eqnarray}
\label{eq:SolutionWindingNumber3}
{\bf s}_{c3}= 2 \tilde \tau_3 \left \{ {\bf b}_{c3} \times {\bf s}_0 + 2 \tau_3 {\bf B}_3 \times ( {\bf b}_{c3} \times {\bf s}_0) \right \} \\
{\bf s}_{s3}= 2 \tilde \tau_3 \left \{ {\bf b}_{s3} \times {\bf s}_0 + 2 \tau_3 {\bf B}_3 \times ( {\bf b}_{s3} \times {\bf s}_0) \right \}, \nonumber
\end{eqnarray}
where we have used that ${\bf B}_3 || {\bf s}_0$ and
\begin{equation}
\tilde \tau_3  = \frac{\tau_3}{1+(2 B_3 \tau_3)^2}.
\end{equation}

Plugging the solutions for the anisotropic components~(\ref{eq:SolutionWindingNumber1}) and (\ref{eq:SolutionWindingNumber3}) back into the equation for the isotropic spin density~(\ref{eq:s0}) and using $\alpha=\beta$, i.e., the condition for the PSH state, one finds for the spin density ${\bf S}$, which is given by ${\bf S}=\int \frac{d {\bf k}}{(2 \pi)^2} {\bf s}_0$  at $T=0$ the spin diffusion equation Eq.~(\ref{eq:SpinDiffusionEquation}).

\end{appendix}


\begin{thebibliography}{99}
\bibitem{WindingNumber3Explanation}
This reasoning is based on the equations of appendix~\ref{app:SpinDiffusionEquation}. When the expression for the winding number three part of the spin distribution~(Eq.~(\ref{eq:Sc3Ss3})) is plugged into the equation for the isotropic component (Eq.~(\ref{eq:s0})) one finds that the only term, which generates a finite $S_x$-component is of the form
${\bf b}_{c3} \times ({\bf s}_0 \times ({\bf b}_{c3} \times {\bf s}_0))=-({\bf b}_{c3} \times {\bf s}_0) ({\bf b}_{c3} \cdot {\bf s}_0)$ where ${\bf b}_{c3} \propto \hat e_y$.
\end{thebibliography}
\end{document}